# Spin-1/2 Triangular Lattice with Orbital Degeneracy in a Metallic Oxide $Ag_2NiO_2$

H. Yoshida,[1] Y. Muraoka,[1] T. Sörgel,[2] M. Jansen[2] and Z. Hiroi[1]

[1]*Institute for Solid State Physics, University of Tokyo, Kashiwa, Chiba 277-8581, Japan*
[2]*Max-Planck-Institut für Festköperforschung, Heisenbergstrasse 1, 70569 Stuttgart, Germany*

A novel metallic and magnetic transition metal oxide $Ag_2NiO_2$ is studied by means of resistivity, magnetic susceptibility, specific heat and X-ray diffraction. The crystal structure is characterized by alternating stacking of a $Ni^{3+}O_2$ layer and a $(Ag_2)^+$ layer, the former realizing a spin-1/2 triangular lattice with $e_g$ orbital degeneracy and the latter providing itinerant electrons. It is found that the $NiO_2$ layer exhibits orbital ordering at $T_s$ = 260 K and antiferromagnetic spin ordering at $T_N$ = 56 K. Moreover, a moderately large mass enhancement is found for the itinerant electrons, suggesting a significant contribution from the nearly localized Ni $3d$ state to the Ag $5s$ state that forms a broad band.

Frustration on triangle-based lattices is one of key ingredients for quantum phenomena to be induced in various electron systems with spin, orbital and charge degrees of freedom. It suppresses classical long-range order (LRO) associated with these degrees of freedom and sometimes leads to a "liquid" ground state with finite entropy.[1] Another interesting issue emerging recently is to clarify the role of frustration on itinerant electrons in frustrated lattices. One would expect some influence especially for a strongly correlated electron system, where both charge and spin degrees of freedom may coexist marginally. A good example is found in a mixed valent vanadium spinel compound $LiV_2O_4$ that exhibits unusual heavy-Fermion like behavior.[2,3] It is suggested that orbital degeneracy survives due to frustration on the pyrochlore lattice and gives rise to a large mass enhancement of $3d$ electrons.

$LiNiO_2$ was first assumed to be a candidate for the spin-1/2 triangular antiferromagnet, because it comprises $Ni^{3+}$ ions in the low-spin state ($t_{2g}^6 e_g^1$) with an isotropic $S$ = 1/2, arranged in a triangular lattice which stacks alternatively with a nonmagnetic $Li^+$ layer.[4] However, later studies have revealed that a net magnetic interaction between neighboring $Ni^{3+}$ spins in $LiNiO_2$ is ferromagnetic (FM), not antiferromegnetic (AFM), and thus there is no frustration for spins.[5] In spite of this, $LiNiO_2$ still remains interesting because of the absence of a cooperative Jahn-Teller distortion that is expected for such a system with orbital degeneracy.[6] Experimental results on $LiNiO_2$ are rather controversial. It always suffers from nonstoichiometry: excess Ni ions replace Li ions, which gives rise to a local interlayer FM coupling that competes with an intrinsic AFM coupling.[5,7] It is believed that ideal $LiNiO_2$ would be a quantum spin-orbital liquid without LRO, where frustration prevents orbital ordering as well as spin ordering in spite of dominant FM couplings.[8] It is pointed out, alternatively, that the absence of LRO is due to the excess Ni ions and that ideal $LiNiO_2$ would be unlikely a spin-orbital liquid but a ferro-orbital state with FM spin ordering,[7,9] as in an isostructural and isoelectronic compound $NaNiO_2$. $NaNiO_2$ exhibits a clear sign of orbital ordering at 480 K and in-plane FM ordering below a much lower temperature of 20 K.[10,11] On one hand, similar orbital physics has been already studied in isostructural $LiVO_2$ with a $V^{3+}$ ion having $t_{2g}$ orbital degeneracy, where a spin-singlet state is driven by special orbital ordering.[12] Hence, it is intriguing to study more about orbital frustration or ordering and its relevance to spin ordering in the triangular lattice.

Here we present another $Ni^{3+}$ triangular lattice in a novel silver oxide. $Ag_2NiO_2$ was found in 2002 by Schreyer and Jansen.[13] It crystallizes in the space group $R-3m$ (No. 166) with $a$ = 0.29193 nm and $c$ = 2.4031 nm. The crystal structure is characterized by alternating stacking of a $NiO_2$ slab and a pair of staggered hexagonal silver layers, as depicted in Fig. 1. The former comprises $NiO_6$ octahedra linked by their edges, just as in $LiNiO_2$ and others. Since the nickel ion is in a trivalent state with seven 3$d$ electrons in a low spin state, the $NiO_2$ layer presents a spin-1/2 triangular lattice, as evidenced by magnetic susceptibility measurements.[13,14] On the other hand, the $(Ag_2)^+$ layer is responsible for metallic conductivity because of the quarter-filled, broad Ag 5$s$ band as in a metallic compound $Ag_2F$.[15] Thus, $Ag_2NiO_2$ is considered to be a unique system where one would expect an interesting interplay between itenerant electrons and localized quantum spins on the frustrated lattice. It was reported that the compound exhibits AFM order below $T_N$ = 56 K, where resistivity showed an anomaly.[13,14] In this paper, we demonstrate in more detail the electronic, magnetic and structural properties of $Ag_2NiO_2$ and discuss on the nature of spins and orbitals in the $NiO_2$ layer. It is pointed out that an AFM superexchange interaction and thus in-plane AFM order are possible for a special orbital arrangement. Moreover, it is shown that a moderately large mass enhancement occurs for the itinerant electrons.

A powder sample was prepared by the solid-state reaction of $Ag_2O$ and NiO under high oxygen pressure as reported previously.[13] Resistivity measurements were carried out down to 0.4 K on a compressed pellet by the standard four-probe method in a Quantum Design Physical-Property-Measurements-System (PPMS) equipped with a $^3$He refrigerator. Magnetic susceptibility measurements were performed down to 2 K in a Quantum Design Magnetic-Property-Measurements-System (MPMS). Specific heat was measured by the heat-relaxation method in PPMS. Low-temperature structure was examined down to 5 K by means of powder X-ray diffraction using CuK$\alpha$ radiation. A silicon powder was mixed with a sample powder and used as an internal standard to determine the temperature dependences of lattice parameters by Rietveld analysis.

$Ag_2NiO_2$ shows metallic conductivity without a superconducting transition down to 0.4 K, Fig. 2. Two anomalies are discernible at $T_s$ = 260 K and $T_N$ = 56 K; a small hump at $T_s$ and a sudden decrease below $T_N$, as reported previously.[13] The latter corresponds to the AF order of Ni spins, as evidenced by magnetic susceptibility data shown in Fig. 3. The $\rho$-$T^2$ plot shown in the inset of Fig. 2 indicates $T^2$ behavior with a coefficient $A$ of 0.033 $\mu\Omega$cm/K$^2$ below $T_N$ and approximately with $A$ = 0.011 $\mu\Omega$cm/K$^2$ above $T_N$. This is a clear sign that electron correlations play a role in scattering carriers. The enhancement of $A$ below $T_N$ means that effective carrier mass $m^*$ is increased, because $A$ is proportional to $(m^*)^2$ in the Fermi liquid picture.

Figure 3 shows the magnetic susceptibility $\chi$ of $Ag_2NiO_2$. It exhibits Curie-Weiss (CW) behavior at high temperature and a cusp at $T_N$. Below $T_N$, there is a small thermal hysteresis which is larger for smaller fields. The temperature dependence of inverse $\chi$ measured at $H$ = 1 T is also shown in

Fig. 3. As noted previously,[13] there is another tiny anomaly around 260 K, where the inverse $\chi$ changes its slope apparently. Effective moment $p_{eff}$ and Weiss temperature $\Theta_W$ are determined as $\mu_{eff}$ = 1.98 $\mu_B$ and $\Theta_W$ = -33 K below 260 K, and $\mu_{eff}$ = 1.77 $\mu_B$ and $\Theta_W$ = 10 K above 260 K. These effective moments are close to the spin-only value expected for a low-spin $Ni^{3+}$ ion, which implies that Ni 3$d$ electrons are nearly localized at high temperature. The small deviation from the $S$ = 1/2 value (1.73 $\mu_B$) must come from spin-orbit couplings through the Lande $g$ factor in the expression of $\mu_{eff}$ = $g(S(S+1))^{1/2}$ $\mu_B$; $g$ = 2.04 above 260 K and $g$ = 2.29 below 260 K. This fact strongly suggests that the local environment around a $Ni^{3+}$ ion is nearly spherical above 260 K, while is distorted to have a certain anisotropy below that. Therefore, there must be a Jahn-Teller type structural phase transition at $T_s$ = 260 K. Another important finding is that $\Theta_W$ changes its sign at $T_s$. The high-temperature value of 10 K means a net FM interaction, as usually observed in other related compounds,[5,10] while the low-temperature value of -33 K points to an unusual AFM interaction. This must be also related to the structural transition, as will be discussed later.

Two anomalies at $T_s$ and $T_N$ are also detected in specific heat, Fig. 4, which gives strong evidence of phase transitions with bulk nature. Both of them may be of the second order. A marked result is found in the $C/T$ versus $T^2$ plot shown in the inset of Fig. 4. From the observed linear relation expressed as $C/T = \gamma + \beta T^2$, where $\gamma$ is a Sommerfeld coefficient, we obtain that $\gamma$ = 18.8 mJ/K$^2$ mol and $\beta$ = 0.713 mJ/K$^4$ mol. The Debye temperature $\Theta_D$ is deduced from the $\beta$ value to be 239 K. The $\gamma$ value of 18.8 mJ/K$^2$ mol is notably large compared with those found in other subvalent silver compounds with a broad Ag 5$s$ band. For example, $Ag_2F$ and $Ag_7O_8NO_3$, both of which are superconductors with $T_c$ = 66 mK and 1.0 K, respectively, have $\gamma$ = 0.62 mJ/K$^2$ mol and 7.28 mJ/K$^2$ mol (about 1 mJ/K$^2$ for 1 mol of Ag), respectively.[15,16] Therefore, a large enhancement in $\gamma$ occurs in $Ag_2NiO_2$.

Figure 5 shows the temperature dependence of lattice parameters, where distinct changes both at $T_s$ and $T_N$ are observed, particularly in the $c$ parameter. Below $T_s$ $c$ begins to increase gradually, while $a$ decreases slightly, suggesting that a second-order structural transition sets in. On the other hand, $c$ suddenly drops at $T_N$. Presumably $c$ is enhanced largely and $a$ is reduced slightly in the intermediate temperature range between $T_s$ and $T_N$. The change at $T_N$ seems to be too large only due to the effect of magnetostriction. We detected neither extra reflections nor a signature for symmetry lowering down to 5 K in the present powder XRD experiments. However, a high-resolution or single-crystalline study is required to clarify the details of the structural transitions.

First we would like to discuss on the superexchange interactions in the $NiO_2$ layers. Although a major difference between the previous $LiNiO_2/NaNiO_2$ and the present $Ag_2NiO_2$ is the existence of conduction electrons, a possible carrier-mediated interaction between Ni spins may be smaller than ordinary superexchange interactions via oxide ions. In previous theoretical study it was pointed out that magnetic interactions through 90º Ni-O-Ni bonds in the $NiO_2$ layer are rather weak and always ferromagnetic, independent of orbital states.[9] Certainly, the Weiss temperature is plus in sign and 20-30 K for nearly stoichiometric $LiNiO_2$[5] and 36 K for the orbital-ordered state of $NaNiO_2$.[10]

In order to get a valuable insight into this issue, we have estimated the sign and relative magnitude of superexchange interactions for all the possible combinations of the twofold $e_g$ orbitals in a pair of $NiO_6$ octahedra connected by their edges, in terms of Goodenough's mechanism modified by Kanamori.[17] The two orbitals could be of $d_{3z2-r2}$ ($d_{z2}$)-type and of $d_{x2-y2}$-type, ignoring the small trigonal distortion of a $NiO_6$ octahedron which results in a Ni-O-Ni bond angle of 95.5º in the case of $Ag_2NiO_2$.[13] The cases considered consist of six by six combinations of orbitals and contain 21 independent combinations. It is found that most of them give net FM superexchange interactions, as typically shown in the lower-left part of Fig. 1b, which is actually the ferrodistorsive orbital ordering pattern of $d_{z^2}$ observed in $NaNiO_2$.[11] Superexchange interactions in this case take place via two bridging oxide ions and are obviously FM, because each pathway contains a pair of orthogonal $p_\sigma$ orbitals of an intervening oxide ion. Note that there are two kinds of FM interactions for the Ni triangle shown in Fig. 1b. In some other cases, however, FM and AFM interactions may coexists on two different pathways, with the FM one larger than the AFM one in magnitude. Remarkably, one exception that must cause pure AFM interactions is found for such a special geometry between two $d_{x^2-y^2}$-type orbitals as shown in the lower-right part of Fig. 1b. In the Ni triangle with all $d_{x^2-y^2}$ orbitals two couplings are AFM and the other is FM. The AFM coupling should occur because either of two $p_\sigma$ orbitals is orthogonal to the opposing $d_{x^2-y^2}$ orbital, that is, there are two sets of orthogonal orbitals in the pathway. Thus, such a ferrodistorsive orbital ordering made of all $d_{x^2-y^2}$-type orbitals can make the net magnetic interaction AFM, providing the magnitude of the FM interaction is smaller than twice that of the AFM one.

The structural phase transition found at $T_s$ = 260 K in $Ag_2NiO_2$ must be due to orbital selection and order. The observed change of the g factor from 2.0 to 2.3 across $T_s$ indicates that a $NiO_6$ octahedron is deformed below $T_s$. Thus, $Ag_2NiO_2$ above $T_s$ should represent an ideal situation with perfect $e_g$ orbital degeneracy. By the way, the g factors of $LiNiO_2$ and $NaNiO_2$ are 2.21[5] and 2.14,[10] respectively. The observed small FM value of 10 K for $\Theta_W$ means that net F interactions survive after statistical averaging over all the possible combinations of orbital states as discussed above. It is to be noted that Yamaura et al. measured $\Theta_W$ in a series of $Li_{1-x}Ni_{1+x}O_2$ samples and found $\Theta_W$ -> 10 K as $x$ -> 0,[5] in good agreement with our result: larger values previously reported may be due to excess Ni ions in the Li layer.

The change of $\Theta_W$ to a larger AFM value of -33 K below $T_s$ is meaningful. It implies that a certain orbital ordering occurs, which must preferentially select such a local orbital arrangement as shown in Fig. 1b to lead to a net AFM interaction. Then, it may be reasonable to assume that a perfect ferrodistorsive long-range orbital ordering occurs by selecting one of $d_{x^2-y^2}$-type orbitals, as suggested theoretically for general cases.[9] Such a ferrodistorsive orbital ordering would induce a monoclinic lattice deformation. However, this may be small, compared with that found in $NaNiO_2$ made of a $d_{z^2}$-type orbital. It is interesting to consider what causes the difference in ordering patterns. The driving force of orbital ordering is probably to lift the $e_g$ orbital degeneracy with the maximum gain in magnetic energy and with the minimum loss in lattice energy arising from a cooperative Jahn-Teller deformation. Therefore, the pattern of orbital order must be determined so as to compromise these two factors. The difference between $Ag_2NiO_2$ and $NaNiO_2$ may come from the difference in structural environments of the $NiO_2$ layer.

The unique feature of $Ag_2NiO_2$ is the coexistence of localized spins on $Ni^{3+}$ ions and conduction electrons from the broad Ag 5s band. In general, 5s electron orbitals are expanded in a crystal and tend to form a broad band. Thus, density-of-states at the Fermi level should be small, giving a small $\gamma$ value in specific heat. In fact, similar subvalent silver oxides like $Ag_2F$[15] and $Ag_7O_8NO_3$[16] have $\gamma$ ~ 1 mJ/$K^2$ mol, as mentioned before. Therefore, the observed $\gamma$ value of 18.8 mJ/$K^2$ mol for $Ag_2NiO_2$ is remarkably large, and strongly suggests that there is a significant contribution from Ni 3d states. According to the band structure calculation, an empty upper Hubbard band from the Ni 3d states lies just above the Fermi level.[13] Rather strong couplings between $5s$ and $3d$ electrons have been noticed in the observed large reduction of $\rho$ below $T_N$ (Fig. 2). Nevertheless, if the $3d$ spins and orbitals have been made completely ordered below $T_s$ and $T_N$, no degrees of freedom would be left to be coupled with conduction electrons. This suggests an intriguing possibility that the spin and/or orbital degeneracy of the $Ni^{3+}$ triangular lattice is lifted only partially by the two transitions due to certain quantum fluctuations.

In conclusion, we have studied a novel metallic and magnetic compound $Ag_2NiO_2$, where the $NiO_2$ layer realizes a spin-1/2 triangular lattice with $e_g$ orbital degeneracy and the Ag layer provides conduction electrons. The former shows orbital ordering at $T_s = 260$ K and magnetic ordering at $T_N = 56$ K, while the latter exhibits large mass enhancement. Interesting interplay between frustrated spin/orbitals and itinerant electrons may be expected.

We are grateful to S. Miyahara and K. Penc for helpful discussions. This research is supported by a Grant-in-Aid for Scientific Research on Priority Areas (Invention of Anomalous Quantum Materials) provided by the Ministry of Education, Culture, Sports, Science and Technology, Japan.

Figure captions

FIG. 1. (Color online) Crystal structure of $Ag_2NiO_2$. A perspective view on the left (a) shows the layered structure made of $Ag_2$ and $NiO_2$ layers, and a plan view on the right (b) illustrates the arrangement of $NiO_6$ octahedra in the $NiO_2$ layer. In the lower-left part of (b) an orbital arrangement observed in $NaNiO_2$ is depicted, which comprises all $d_{3z^2-r^2}$ ($d_{z^2}$) orbitals inducing two kinds of ferromagnetic superexchange interactions, F and F'. In the lower-right part of (b) an alternative orbital arrangement possibly realized in $Ag_2NiO_2$ below 260 K is shown, where only the $d_{x^2-y^2}$ orbital is selected for a Ni trimer giving two antiferromagnetic and one ferromagnetic interactions.

FIG. 2. (Color online) Resistivity $\rho$ of $Ag_2NiO_2$ measured on a compressed pellet. $T_s$ and $T_N$ are the temperarures of structural and magnetic transitions. Inset shows a $\rho$-$T^2$ plot below about 90 K. The lines are guides to the eye.

FIG. 3. (Color online) Temperature dependences of magnetic susceptibility $\chi$ on the left axis and inverse susceptibility $1/\chi$ on the right axis measured at $H = 1$ T . In the former open marks represent data measured on heating after zero-field cooling, and solid marks are from the following cooling experiments. The two solid lines for the latter show Curie-Weiss plots above and below $T_s = 260$ K.

FIG. 4. (Color online) Specific heat $C$ of $Ag_2NiO_2$ measured on cooling from room temperature showing two anomalies at $T_s$ and $T_N$ . Inset shows specific heat divided by temperature plotted against $T^2$ showing a linear change at low temperature below 11 K. The solid line is a fit to the equation, $C/T = \gamma + \beta T^2$.

FIG. 5. (Color online) Temperature dependences of lattice constants $a$ and $c$. Error bars on the data are from Rietveld analyses.

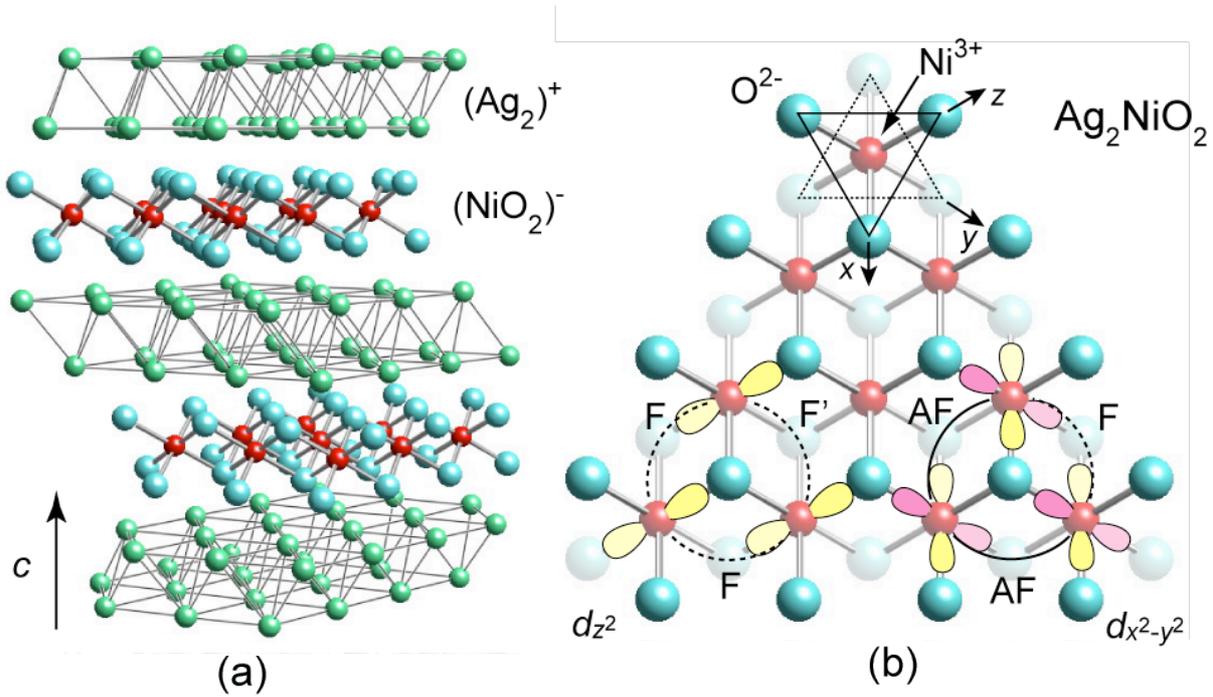

Fig. 1

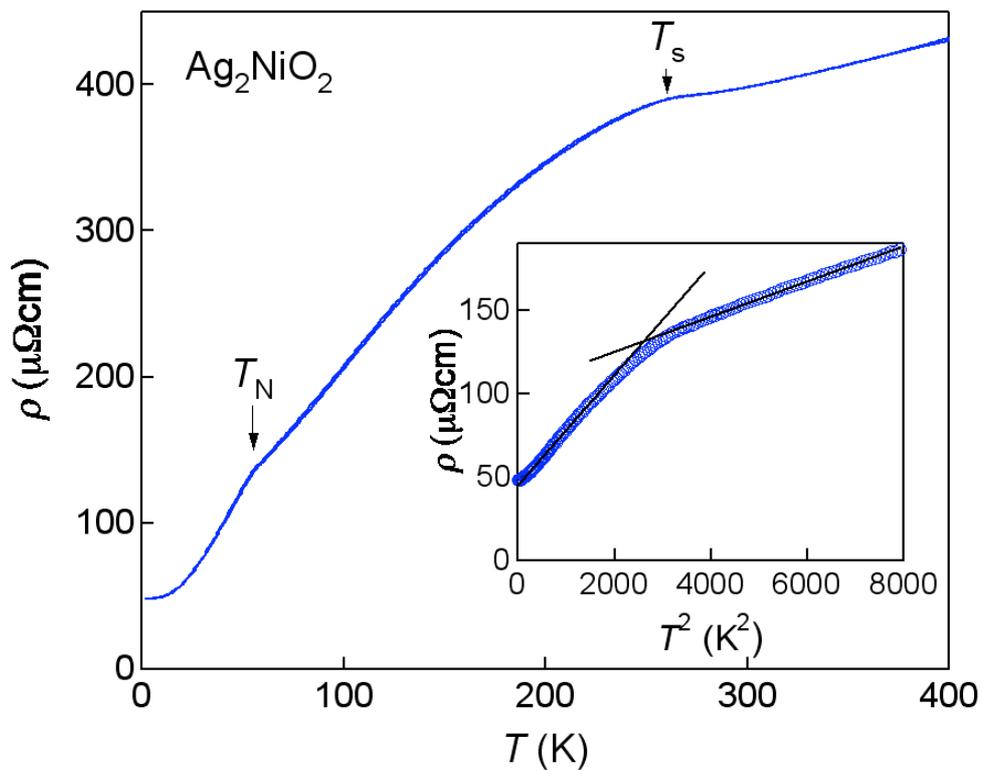

Fig. 2

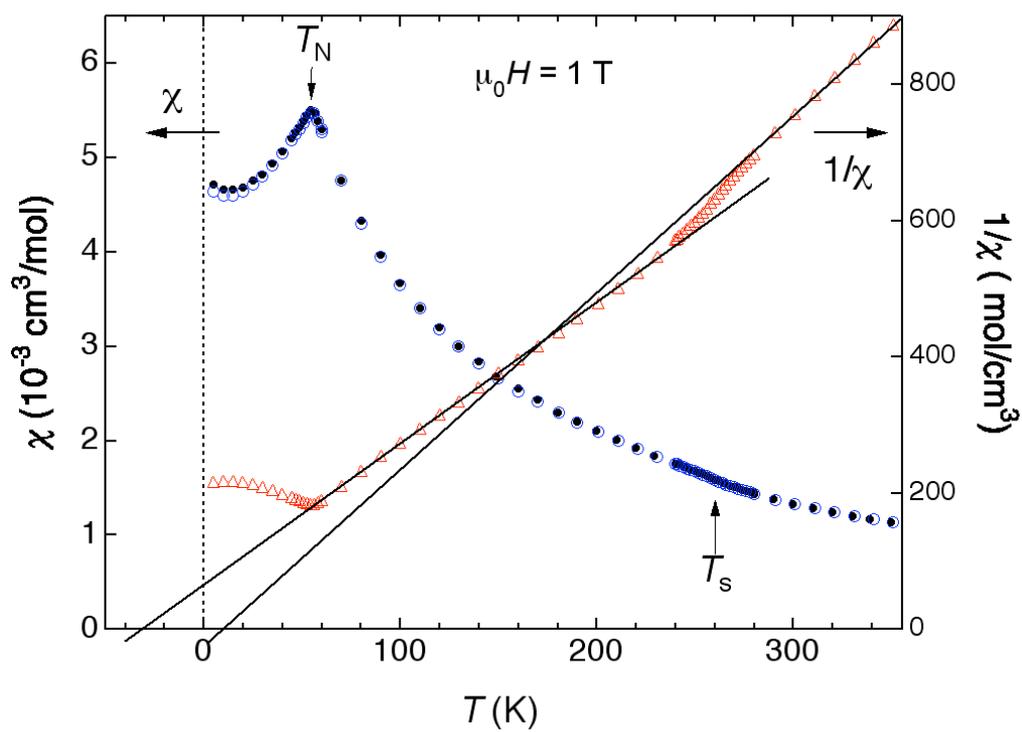

Fig. 3

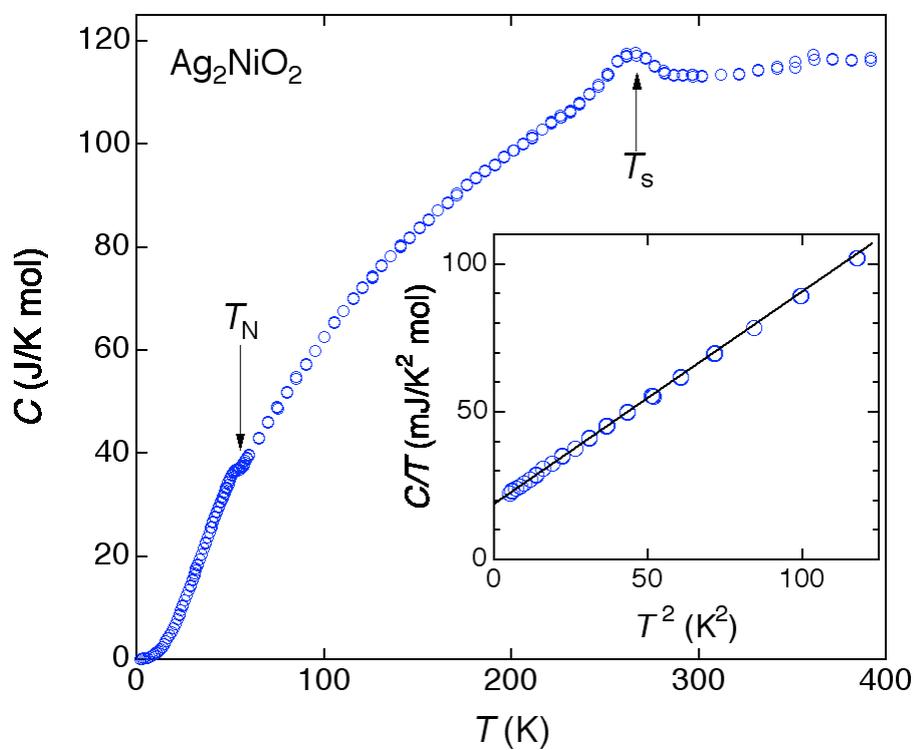

Fig. 4

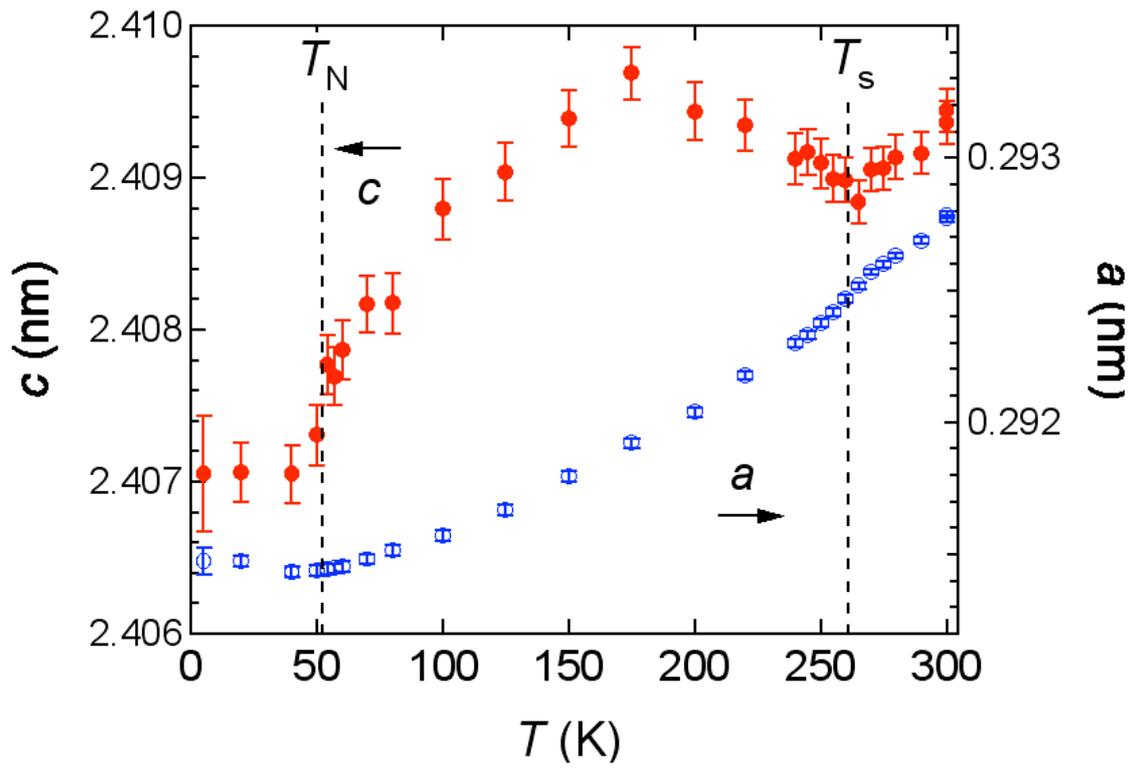

Fig. 5